\documentstyle[editedvolume,psfig]{crckapb}

\begin{opening}
\title{The Square Kilometer Array Interferometer}
\author{Robert Braun}
\institute{Netherlands Foundation for Research in Astronomy}
\end{opening}
\runningtitle{The Square Kilometer Array Interferometer}
\begin{document}
\vskip -1cm
\centerline{\it To appear in: ``The Westerbork Synthesis Telescope,}
\centerline{\it \quad \quad It's 25$^{th}$ Anniversary and Beyond''}
\centerline{\it Eds. E.Raimond \& R.Genee (Kluwer, Dordrecht)}

\section{Introduction} The rapid pace of technological development
within our civilization has fueled an explosion in our capability to
study nature, which in turn has spurred the pace of further technology
development and so on \dots This feed-back cycle leads naturally to an
exponential growth of our ability to study nature as measured by almost
every imaginable criterion. Whether we chart the evolution of particle
accelerator beam energies, the resolution of microscopes, or the
performance of computers we see an exponential increase in performance
which extends back in time to at least 1930 for many applications. The
improvement in performance over the past 65 years can be in excess of
nine orders of magnitude, while the facility cost (in adjusted terms)
has not necessarily increased dramatically.

This exponential improvement in  performance is illustrated for the
case of particle accelerator beam  energies in Figure~1. If we look in
detail at the time evolution of a capability, we notice that it is
built up of numerous sub-tracks, each due to a particular technology.
The sub-tracks tend to saturate at a specific performance when that
technology no longer allows increased cost-effectiveness. This
saturation seems to reflect the fact that there is a `Basic Facility
Cost' (BFC) in most branches of research which `Society' is willing to
invest in the future. Staying on the exponential growth curve demands
the continuous development of new technologies which allows order of
magnitude performance improvements at a fixed BFC. If, for whatever
reason, a branch of research strays significantly below the exponential
growth curve, then the rate of discovery may be expected to fall off
dramatically. Interest in the field declines and the most talented
researchers generally choose to invest their energies in alternate
research directions. Conversely, if one can take advantage of an
emerging technology to re-join or overtake the exponential growth curve
at a fixed BFC, the discovery rate will surge and the field will
blossom.

\begin{figure}[htb]
\caption{Livingstone curves charting
the evolution of particle accelarator beam energies}
\end{figure}

The same trends are apparent in the evolution of our capabilities in
observational (and computational) astronomy. The instantaneous
sensitivity and angular resolution of observations in virtually all
frequency bands has been improving exponentially with time. The
frequency bands least accessible from the earth's surface, X-rays and
the far infrared have had to wait for appropriate detection
technologies and space-based observing platforms. One implication is
that the BFC is rather different for the different frequency bands,
with a particular distinction between those demanding space based,
versus those possible from earth based platforms. The fact that much
higher BFC's are accepted for some applications illustrates that there
are political and economic as well as purely scientific considerations.
However, we have now reached the evolutionary stage where all
electromagnetic frequency bands and many flavours of particle
astrophysics have been opened up to exploration. The capabilities in
each band are now growing along their own tracks. The capability
doubling time along a track, and in some sense the ``discovery rate'',
is determined not so much by the ``Basic Facility Cost'' associated
with that band (unless it is highly time variable), but by whether or
not there is a continued application of new technologies.

\section{Decimetric Radio Astronomy in Perspective} Let us now look
more specifically at the evolution of sensitivity at decimetric radio
wavelengths. In Figure~2 we plotted the continuum sensitivity after one
minute of integration for many radio telescopes as they became
available over the last 60 years. Since upgraded receiver systems have
often been added to existing facilities after construction, these are
also indicated in the figure in a number of cases. An exponential
improvement in system performance over at least 6 orders of magnitude
is apparent between about 1940 and 1980. Instruments like the WSRT and
the VLA have become available on a schedule which maintained a high
rate of discovery. The Arecibo telescope stands out as a major leap in
sensitivity performance at a relatively early date. This example serves
to illustrate that the single parameter we've chosen to examine in
Figure~2, the continuum sensitivity, doesn't necessarily tell the
entire story. Additional parameters, like sky coverage and survey speed
also play a significant role in defining the total system performance.
Even so, a trend seems to have developed in the period 1980 to 1995
toward significant saturation of performance obtained with traditional
radio telescope technology.

\begin{figure}[htb]
\caption{
The time evolution of radio telescope sensitivity. The continuum
sensitivity after one minute of integration is indicated for a number of
radio telescopes as they became available. Solid lines indicate up-grade
paths of particular instruments. }
\end{figure}

This saturation may already be paired with a declining discovery rate
and migration of researchers. Such a migration is stimulated by the
prospect of access to exponential growth curves (and high discovery
rates) in other frequency bands, such as the new generation of 8~meter
class optical and near infrared telescopes or the second generation of
millimeter arrays.

During the period 1980 to 1995, decimetric receiver technology has
advanced to the point where receiver losses no longer dominate the
system performance. Indeed, at these wavelengths, near state-of-the-art
system performance can now be achieved with highly integrated
components operating at room temperature and costing only a few Dutch
guilders each. Major improvements to cost-effectiveness have been
hampered by the high fixed costs of the large mechanical structures in
traditional radio telescope designs.

The time is therefore now ripe to pursue a radically different
technology in the decimetric band; one in which the economies of scale
are applied to high performance, yet extremely low cost amplifiers and
digital electronics, while the dependence on large mechanical components
is minimized. The successful realization of such a breakthrough, with a
one to two order of magnitude surge in performance, would have an
enormous impact on a wide range of astrophysical studies.

\section{Scientific drivers for a major new instrument}
\paragraph{Cosmology} \ \ \\ Let us briefly consider a few of the
research areas in which we can foresee a major impact from
substantially improved performance in the decimetric band. More
detailed discussions of particular topics can be found elsewhere in
this volume. Perhaps the most basic capability we wish to acquire is
the ability to detect and study the gaseous content of massive
condensations throughout the cosmos. The epoch when our own part of the
universe became transparent to radiation corresponds to a time when
about 90 percent of all baryonic matter was in the form of neutral
atomic hydrogen. Many poorly understood processes have modified the
state of matter at subsequent times. While condensations accumulated in
regions of excess density, much of the intervening space was probably
re-ionized by the intense radiation of the first massive stars and
QSO's. Some galaxies apparently contain stars mostly from such early
epochs. Galaxies like our own contain only a small population of old
stars dating to the first epoch of star formation, while the majority
of stars were formed in fits and starts at much later times. Other
galaxies seem to have led a much less eventful life and have had very
little star formation activity to this day. In one way or another,
during the last 10 billion years the neutral gas content of the
universe has declined to only a few percent of the baryonic matter. How
exactly did this happen? When did the lights really turn on in the
universe?

The 21-cm line of neutral hydrogen remains the most valuable tracer of
neutral gaseous mass in astrophysics. Even though neutral gas becomes
predominantly molecular at high densities, the 21-cm line emission of
the atomic component allows the total gaseous mass to be estimated to
better than about a factor of two even in the most extreme cases. This
is in marked contrast to, for example,  the luminosity of carbon
monoxide emission lines originating in the molecular component. For the
same total gaseous mass, these emission lines can vary in luminosity
over a factor of about 10,000 depending on the abundance of heavy
elements (Carbon and Oxygen) and the intensity of illumination to which
they are subjected.

The availability of such a potent tracer of gaseous mass in the
universe as the 21-cm emission line, makes it particularly frustrating
that current instrumentation allows us to probe our own backyard
only(in a cosmological sense). Long integrations with the most
sensitive radio telescopes now available have only enabled us to reach
out to recession velocities of about 30,000~km~s$^{-1}$ corresponding
to look-back times of about one billion years. While this may seem
substantial, it is only about 10 percent of the age of the universe and
only 20 percent of the age of our planet. Our situation is much like
describing an iceberg based solely on what we see protruding from the
water. So much remains hidden from view, including the most exciting
phases when most of the activity took place.

Roughly speaking, we wish to probe distances which are about ten times
as great and therefore require about 100 times the sensitivity of a
100~meter diameter filled aperture. This basic requirement on
sensitivity translates into a collecting area of about one square
kilometer, which has led to the most commonly used name for such an
instrument, ``The Square Kilometer Array Interferometer'' (or SKAI).

\paragraph{Pulsars}\ \ \\ While it is cosmology that drives one to
consider a telescope of such substantial collecting area, other areas
of physics may be expected to benefit dramatically from improved
decimetric sensitivity. For example, the pulsed radio emission from the
most rapidly rotating, weakly magnetized, neutron stars (millisecond
pulsars) represents the most accurate time-keeping device yet
discovered. Nature has been kind enough to distribute these
hyper-accurate clocks throughout our own and other galaxies. This puts
us in a position to actually carry out many of the thought experiments,
considered by the great theoretical physicists of the 20th century, and
others not previously considered. We can test the predictions of
General Relativity in detail through detection and study of the decay
of the most compact binary orbits. We can search for the presence of
black holes or planetary companions via their periodic orbital
perturbations. We can form a gravity-wave telescope employing a grid of
reference pulsars distributed over large separations on the sky. The
only catch in all of these endeavours is the intrinsic rarity of the
millisecond pulsars themselves as well as that of the particularly
interesting circumstances, like the occurrence (and lifetime) of
extremely compact binary orbits and black-hole companions. Nature has
been kind, but not infinitely generous. The only practical means of
identifying and studying those rare cases with a particular utility
lies in a significant expansion of the detection volume. Clearly, the
greater the improvement in sensitivity, the greater will be the impact
on this branch of astrophysics. More specifically, overcoming the
current small number statistics will require a 1000-fold increase in
detection volume, requiring a 100-fold improvement in sensitivity.

\paragraph{Star formation and galactic nuclear activity}\ \ \\ A third
broad category of research opportunity lies in the better utilization
of non-thermal radio continuum emission as a long-lived probe of both
the star-formation history in galaxies and their nuclear activity.
During the last decade, studies utilizing the very successful InfraRed
Astronomy Satellite (IRAS) have revealed the surprising fact that a
rather constant fraction of the energy, released during the lifetime of
massive stars (those with more than about seven times the mass of the
Sun), is converted into the synchrotron radiation visible at decimetric
wavelengths. With sufficient sensitivity and resolution, this
thermometer of the star-formation activity can be applied to galactic
systems at arbitrarily large distances. A major advantage of this
tracer over others, like the optical recombination lines of ionized
hydrogen, is that it is unaffected by the presence of intervening dust
particles, which leads to substantial obscuration of the optical and
ultraviolet light. In addition, a well-designed decimetric instrument
could routinely produce images with a spatial resolution of a fraction
of an arcsecond; a resolution which at optical wavelengths can only be
routinely achieved with space based telescopes, like the Hubble Space
Telescope.

Such high resolution also makes it possible to distinguish the
synchro\-tron emission of the star forming galactic disks from the
energetic phenomena which occur in the most central regions of many,
perhaps all, galaxies. At some phase in the evolution of galactic
systems, they seem to acquire a massive black hole at their center.
These objects often make their presence known by ejecting narrow beams
of energetic particles which can travel distances comparable to those
separating galaxies before dissipating their energy in enormous lobes.
Understanding the creation and evolution of the ``central monsters''
which populate galaxies demands the highest possible angular resolution
while retaining the highest possible sensitivity.

The technique of Very Long Baseline Interferometry (VLBI) can be
brought to bear to achieve angular resolutions of milliarcseconds or
even microarcseconds when orbiting or moon-based telescopes are
available. Although the resolution of space VLBI is particularly
attractive, applications are likely to remain severely limited by the
low sensitivity of the radio telescopes which we can afford to place in
earth orbit. However, since the sensitivity of each telescope
combination in a VLBI observation depends on the product of the two
individual telescope sensitivities, the limitations of the space-based
antenna can be compensated to a substantial degree by construction of
an appropriately more sensitive telescope on the ground. For example,
the combination of a 10~meter orbiting telescope with a square
kilometer telescope on the ground would have the same, high sensitivity
as the combination of two 100~meter diameter telescopes.

\section{Technical specifications} Although it remains essential to
consider the many branches of current research and to predict how they
might be effected by a significant improvement in our capabilities,
history has taught us that our imagination consistently falls short of
the surprises that lie in store for us when a large step in
instrumental capability is achieved. Much of the beauty of science lies
in the fact that nature has managed to provide us with an onion-skin
parameter space, where unexpected discoveries accompany each new layer
to which we penetrate. We must strive to make the most general possible
improvements to our capabilities to maximize the cross-section for
unexpected discoveries. A particular challenge lies in balancing the
breadth of a new instrument against its depth in particular areas,
while keeping within the constraints of the `Basic Facility Cost'.

Since we are strongly motivated by the desire to utilize the
red-shifted 21-cm line to probe gaseous concentrations throughout the
universe, we are led to a natural constraint on the frequency coverage
of the required instrument; namely that it extends continuously from
about 200~MHz to 1420~MHz. This same frequency range is also optimum
for the study of pulsars since their emission spectra rise so
dramatically toward lower frequencies, at least until propagation
effects introduce an effective low frequency cut-off. The (red-shifted)
emission lines of the hydroxyl molecule (OH), with rest frequencies
between 1612 and 1721~MHz, also constitute a worthwhile design goal.
The many continuum applications of such an instrument, in particular
those involving Very Long Baseline Interferometry (VLBI), benefit from
the highest possible upper frequency bound. This is motivated by
achieving high spatial resolution by obtaining the longest possible
baseline lengths (as measured in multiples of the observing wavelength)
as well as circumventing propagation effects both within the emitting
sources as well as along the line-of-sight through our Galaxy.

Consideration of the diverse scientific drivers suggests a design goal
for the frequency range of about 200~--~2000~MHz. An important
difference of this specification with that of existing instruments is
that there be access to large contiguous segments of the frequency
range. Traditionally, only a small number of fairly narrow bands with
relative bandwidths of 10 or 20 percent have been available in this
range. Satisfying this requirement presents at least two new technical
challenges. The first is the design of exceptionally broad-band yet
high performance receiving elements, or alternately, a large suite of
easily accessible narrow-band elements. The second is devising methods
of successfully observing in portions of the decimetric spectrum which
are being actively utilized by other services.

Most of the frequency range, 200~--~2000~MHz, has in fact been
allocated to a variety of active applications including military and
civilian communications, television transmission and navigation aids.
Transmission levels are typically so strong, relative to the extremely
faint levels of radiation we wish to observe, that single element radio
astronomy (as with the Arecibo telescope) is precluded over large
portions of this range, even when the interfering signal has not
saturated the receiver electronics. The reason for this is that a
single element radio telescope does not have the ability to determine
from which direction the signal from a transient source has arrived.
When multi-element radio telescopes are used to sample the correlation
of the incident radiation field it becomes possible to produce crude
images of the entire sky with the data obtained from arbitrarily short
time intervals. If a sufficiently large number of well-distributed
correlation samples are obtained in each short time interval, then
interfering sources can be identified, modeled and successfully removed
from the data. In addition, when the element separation becomes a
non-negligible fraction of the distance to an interfering source then
there is a significant reduction in the correlated interfering power.
Multi-element arrays, like the WSRT and the VLA, are much less
susceptible to data loss due to external interference than are single
element telescopes. However, with only a linear array of fourteen or a
`Y' arrange ment of twenty-seven elements to correlate, the ability to
successfully model transient interfering sources remains limited. This
ability would be substantially enhanced with a fully two dimensional
arrangement of the elements as well as the largest possible number of
instantaneous correlations. A minimum requirement is likely to be for
about 32 correlated elements, although this is an area where ongoing
research will be directed.

In a similar vein, when the individual elements of a multi-element
telescope are themselves generated by the complex addition of many
smaller receiving elements, the response function of that sub-array may
be actively tailored to have a desired form. With a sufficiently large
number of receiving elements (10,000 or more) it becomes possible to
create either an ultra-low mean side-lobe level or extremely deep
response nulls in desired directions, like those of interfering
sources. In this way we could prevent the power of interfering sources
from entering the correlation data at all.

\paragraph{Design goals}\ \ \\ The preceding discussion has illustrated
some of the design considerations for a next generation decimetric
facility, namely:

\noindent $\bullet$ a total collecting area of about 1~km$^2$,

\noindent $\bullet$ distribution over at least 32 elements,

\noindent $\bullet$ a preference for forming even these elements from
many smaller ones,

\noindent $\bullet$ a frequency range of about 200~--~2000~MHz.

Some important additional design goals are the following. The angular
resolution should be as high as possible while maintaining the
necessary sensitivity to low surface brightness emission. In this way,
the morphology of distant gas concentrations can be imaged rather than
only detected as an unresolved feature. On the other hand, the thermal
nature of the 21-cm line emission, places an upper limit on the surface
brightness with which it radiates. The implication is that the angular
resolution must be well-matched to the instrument sensitivity. In
practise, with the sensitivity we are considering, an angular
resolution of about one arc-second is permitted, corresponding to array
dimensions of about 50~km. Such an area is also near the maximum size
that can be coherently summed readily, such as would be done for
observations of pulsars or for participation in a VLBI observation. The
reason for this constraint is that the line-of-sight propagation or
``ionospheric weather'' becomes significantly different for regions
separated by more than a few ten's of kilometers. On the other hand,
the wider bandwidths and higher emission brightnesses available to
continuum observations imply that these would benefit from even higher
angular resolution. One way to satisfy these conflicting demands is to
place a large fraction (say 80 percent) of the collecting area of the
instrument within a region of about 50~km and to distribute the
remaining fraction over a region of about 300~km diameter. In this way,
the applications most limited by low surface brightnesses would retain
most of their sensitivity, while other applications could utilize
angular resolutions as high as about one tenth of an arcsecond.

A very schematic representation of what the SKAI might look like is
given in Figure~3. The shaded circles indicate the position of the unit
telescopes of the array. Note that the size of each circle has been
greatly exaggerated to allow it to be seen on the scale of the
illustration. A rather dense ring-like concentration of the telescopes
over a region of about 50~km extent determines the beam size for which
the brightness sensitivity is optimized (about 1~arcsecond at a
frequency of 1420~MHz), while the additional elements, distributed over
a 300~km extent, would make sub-arcsecond imaging possible over the
entire frequency range of 200~--~2000~MHz.

\begin{figure}[htb]
\caption{
Schematic configuration of SKAI. Note that the unit telescopes are not
depicted to scale. }
\end{figure}

\section{Instrumental capabilities} The capabilities of this telescope
for observations of moderately red-shifted neutral hydrogen emission
are illustrated in Figure~4. Simulated observations are presented of
the nearby spiral galaxy M101 after rescaling actual data appropriately
for distances corresponding to red-shifts of 0.2, 0.45 and 0.9, or
look-back times of 2, 4 and 6 billion years. Realistic measurement
noise has been included. Images of the brightest detected emission are
shown in the left hand panels, while the corresponding relative
velocities are shown on the right. The velocities are only shown where
the emission brightness exceeds four times the noise level. It becomes
clear that detailed kinematic studies of normal galaxies with
sub-arcsecond resolution would be possible out to red-shifts greater
than 0.2, while cruder kinematic studies would still be possible out to
red-shifts of about 1.

\begin{figure}[htb]
\caption{
Simulated SKAI observations of M101 as it would appear at red-shifts of
0.2, 0.45 and 0.9. Peak observed brightnesses are shown in the left hand
panels and corresponding velocity fields on the right. The assumed
integration time is indicated above each panel. }
\end{figure}

It should be borne in mind that these are lower limits to what will be
possible, since the {\it current} gas mass of M101 has been assumed in
the simulations. All indications are that there has been a strong
evolution of gas mass into stellar mass during this time interval. In
particular, deep optical and near-infrared observations have revealed a
population of luminous blue galaxies which densely cover the sky beyond
red-shifts of about 0.4.

Another important point to note is that although the field size
depicted in each panel of Figure~4 is only 230~kpc on a side, or about
one quarter of the typical distance between galaxies in a group, the
actual field of view of the strawman design will be about 1.5, 3 and
5~Mpc in the three cases pictured, while the observing bandwidth will
typically probe a cylindrical volume which is about 200~Mpc deep. This
means that every observation is likely to allow kinematic study of a
large sample of galaxies rather than only a single target.

The capability of the SKAI for simultaneously studying large numbers of
gaseous objects in the early universe is illustrated well in Figure~5.
This figure presents the result of a super-computer simulation in which
structure formation has been traced from the tiny fluctuations thought
to be present at a red-shift of 50 to the depicted situation at
red-shift 2. The grey-scale distribution is proportional to the
logarithm of the mass of neutral hydrogen predicted to lie within each
beam area of the telescope. The dark ``beads'' on the fainter
``strings'' in the figure are proto-galaxy condensations with typical
masses between about 10$^6$ and 10$^{11}$ solar masses; comparable to
the range of galaxy masses which we now see. The white contours
overlaid on the dark proto-galaxy ``beads'' show the detection
threshold of the telescope after an integration time of about 10 weeks.
Hundreds of individual galaxies should be visible in each telescope
field, which has a size comparable to that illustrated in the figure.
An important difference is that while the line-of-sight depth of the
simulated volume in the figure corresponds to only about 8~Mpc, the
typical line-of-sight depth of an actual observation will be about
twenty times greater, so that literally thousands of proto-galaxies
will be studied in each such observation. While it may seem excessive
to study thousands of objects at one epoch in the distant past, this is
exactly what is required to answer some of the most fundamental
questions concerning the nature of our universe. Within the actual
clustering of such distant sources is coded the answer to the question
of whether our universe is open or closed as well as insight into the
nature of both the visible and the dark matter.

\begin{figure}[htb]
\caption{
Simulated HI emission at z = 2 with SKAI detections overlaid. The
linear grey-scale indicates the predicted peak brightness of HI
emission in a 22.2/(1+z) Mpc cube and extends from $log(M_\odot/Beam)$
= 1.7 -- 10.8. The single white contour at $log(M_\odot/Beam)$ = 9.22
is the 5$\sigma$ SKAI detection level after a 1600 hour integration. }
\end{figure}

A different view of how the sky will appear to the SKAI is given in
Figure~6. A deep integration obtained with the Hubble Space Telescope
has been calibrated and re-scaled to simulate the predicted continuum
emission expected from normal galactic disks at a frequency of
1400~MHz. The continuum sensitivity of the array corresponds to about
0.1 brightness units (micro-Janskys per beam) while the saturation
level in the image corresponds to 1.0 of these brightness units. Actual
radio continuum images will differ from this simple simulation in a
number of important ways. Firstly, the field of view of each of the
radio images will have about 100 times the area of the Hubble Space
Telescope images. Secondly, they will not suffer from dust extinction.
Although this is a fairly obvious optical effect in the inclined galaxy
seen in the lower right hand corner of the image, an unknown degree of
internal dust extinction affects the light and structure of all the
galaxies in the image. And finally, the bright non-thermal radio
emission of a compact active nucleus and its associated high-velocity
outflow will be seen within a few percent of the galaxies. The
sensitivity to detect such components and the resolution to distinguish
them from their unobscured host galaxies may well provide important
insights into the phenomenon of nuclear activity.

\begin{figure}[htb]
\caption{Simulated deep radio continuum image. The grey-scale extends
from $-0.1$ to 1.0 $\mu$Jansky per beam.}
\end{figure}

\section{Telescope concepts} The most cost-effective concept for the
unit telescopes of the array must still be determined, but it may
resemble one of the panels in Figure~7. In the top panel, the
adaptively phased array concept is illustrated. Rather than employing
any concentration of the incoming radio waves with a reflector, the
energy is detected directly with many matched receiving elements which
are individually amplified. The amplified energy of all of the
receiving elements in a unit telescope is summed together with a
carefully chosen set of complex weights. These weights can be used to
form a very clean telescope beam, with a low sidelobe level and with
deep nulls in particular directions, like those of interfering sources.
The weights are also used to electronically re-direct the beam on the
sky, so as to allow observation of sources as far as 45 degrees from
the zenith and tracking them for an extended period.

\begin{figure}[htb]
\caption{Possible element concepts for the SKAI. }
\end{figure}

One of the most attractive aspects of such a concept is the complete
absence of mechanical components. Pointing and tracking is achieved
completely electronically, so that with appropriate beam-forming
electronics any desired number of independent, simultaneous beams can
be placed on the sky. These beams might sometimes be used in concert to
simultaneously view a large connected region of the sky, or they might
be used to carry out completely independent observing programs for a
number of researchers at once. One challenge associated with this
concept is in the design of receiving elements to have both good
efficiency and large bandwidth. It seems likely that a small number of
interleaved sets of elements, each optimized for a different center
frequency, will be necessary to adequately cover the frequency range of
interest. In addition, the large numbers of receiving elements required
at the higher frequencies (about 1~million) imposes strong constraints
on the allowed cost of each amplifier, the data transmission and the
electronic combination. Development of extremely low-cost, yet
high-performance components for adaptive beam formation is an area of
very active current research with strong commercial incentives for
finding satisfactory solutions.

Another concept for the unit telescopes is illustrated in the center
panel of Figure~7. In this case, a modest amount of concentration of
the incoming energy is obtained with a close-packed array of small
parabolic reflectors. The energy is then detected, amplified and
combined with appropriate weights just as discussed above. Since the
incoming energy has already been concentrated, a much smaller number of
components (about 1000) is necessary for signal amplification,
transmission and combination. The price paid for this electronic
simplicity is the added mechanical complexity of a pointing and
tracking mechanism for the paraboloids. Many of the benefits of the
fully adaptive array concept still apply, including the ability to form
multiple observing beams individually tailored, when necessary, to have
low sidelobes and deep nulls in the directions of interfering sources.
However, the capability of allowing completely independent observing
programs concurrently would be lost, since all simultaneous beams would
be constrained to lie in the field of view of an individual paraboloid.
Fabrication of low cost, low maintenance paraboloids is now practical
with fiber-reinforced epoxy technology. The major challenge to overcome
with this concept is the design of a similarly low-cost and
low-maintenance drive system. On-going advances in the area of
lubrication-free mechanical systems may offer a possibility for
over-coming this challenge.

At the other physical extreme is the concept depicted in the bottom
panel of  Figure~7. A stationary spherical reflector is employed to
concentrate incoming energy at an approximate focus along the spherical
surface near one half its radius of curvature. Further energy
concentration is accomplished with either an additional, specially
shaped, reflector or directly with a large array of receiving elements.
The receiving structure must be mechanically yositioned along the
spherical focal surface to enable pointing and tracking of sources on
the sky. Designing an effective, low-cost mechanism for the support and
tracking of the receiving structure probably represents the greatest
engineering challenge for this concept. Although multiple observing
beams could be generated with this approach, they would be limited to
lying very near each other on the sky. Possibilities for interference
rejection by beam tailoring would only exist in the case where a
sufficiently large number (perhaps 1000) of receiving elements were
employed. One prerequisite for such a concept to be cost-effective is
the availability of appropriate terrain for telescope construction. The
Arecibo telescope, which is of a similar design, was constructed in a
region of so-called Karst formations. When regions of sedimentary
limestone deposits are subjected to intensive rainfall over geological
times, hemispherical depressions often develop as a result of the roof
collapse of dissolved subterranean cavities. Such formations can now be
found in a number of tropical and semi-tropical regions, particularly
in South and Central America as well as in South-East Asia. An
extensive site survey conducted in China based on satellite, aerial and
ground-based data has identified a potential site for an array of
spherical reflector telescopes within the Guizhou province in
South-Central China. In this sparsely populated region, hundreds of
hemispherical depressions of appropriate dimensions are found in an
area which is several hundred kilometers in diameter.

Several other telescope concepts are currently being explored,
including the use of reflect-array technology and extremely long focal
length adaptive paraboloids. Further technical research in the coming
few years should allow the most promising concepts to be identified and
refined. Prototypes based on several different concepts are likely to
be built in the Nether lands and in China. If at least one truly
cost-effective technology emerges from this develop ment phase, then
funding will be sought for a construction start early in the next
century. In this way we can look forward to opening the door to a new
era of discovery.

\end{document}